\documentclass[prl,superscriptaddress,tightenlines,twocolumn,showpacs,floatfix]{revtex4-1}
\usepackage{amsmath}
\usepackage{amsfonts}
\usepackage{amssymb}
\usepackage{color}
\usepackage{graphicx,float}
\usepackage{chemarr}

		\newcommand{\ro}[0]{\rho_{\mathrm{1}}}	
		\newcommand{\rt}[0]{\rho_{\mathrm{2}}}
		
		\newcommand{\rtnd}[0]{\rho_{\mathrm{2}}^\mathrm{nd}}

		\newcommand{\re}[0]{\rho_{\mathrm{tot}}}

		\newcommand{\kp}[0]{k_+}
		\newcommand{\km}[0]{k_-}
		\newcommand{\kmeff}[0]{k_-}
		\newcommand{\tkmeff}[0]{\tilde{k}_-}
		\newcommand{\kpeff}[0]{k_+}

		\newcommand{\Kd}[0]{K_\mathrm{d}}
		\newcommand{\Deff}[0]{D_\mathrm{eff}}
		\newcommand{\Vdis}[0]{\boldsymbol{V}_\mathrm{dis}}
		
		\newcommand{\tD}[0]{\tilde{D}_2}

\begin{document}

\title{Cooperatively enhanced reactivity and `stabilitaxis' of dissociating oligomeric proteins}
\author{Jaime Agudo-Canalejo}
%\email{jaime.agudo@ds.mpg.de}
\affiliation{Max Planck Institute for Dynamics and Self-Organization (MPIDS), D-37077 G\"ottingen, Germany}
\affiliation{Rudolf Peierls Centre for Theoretical Physics, University of Oxford, Oxford OX1 3PU, United Kingdom}
\affiliation{Department of Chemistry, The Pennsylvania State University, University Park, Pennsylvania 16802, USA}
\author{Pierre Illien}
\affiliation{Sorbonne Universit\'e, CNRS, Laboratoire PHENIX, UMR CNRS 8234, 75005 Paris, France}
\author{Ramin Golestanian}
\email{ramin.golestanian@ds.mpg.de}
\affiliation{Max Planck Institute for Dynamics and Self-Organization (MPIDS), D-37077 G\"ottingen, Germany}
\affiliation{Rudolf Peierls Centre for Theoretical Physics, University of Oxford, Oxford OX1 3PU, United Kingdom}

\begin{abstract}

Many functional units in biology, such as enzymes or molecular motors, are composed of several subunits that can reversibly assemble and disassemble.
This includes oligomeric proteins composed of several smaller monomers, as well as protein complexes assembled from a few proteins.
By studying the generic spatial transport properties of such proteins, we investigate here whether their ability to reversibly associate and dissociate may confer them a functional advantage with respect to non-dissociating proteins.
In uniform environments with position-independent association-dissociation, we find that enhanced diffusion in the monomeric state coupled to reassociation into the functional oligomeric form leads to enhanced reactivity with localized targets.
In non-uniform environments with position-dependent association-dissociation, caused e.g.~by spatial gradients of an inhibiting chemical, we find that dissociating proteins generically tend to accumulate in regions where they are most stable, a process that we term \emph{stabilitaxis}.

\end{abstract}

\date{\today}

\maketitle

It has become increasingly clear in recent years that, in order to fully understand intracellular reaction pathways, it is not sufficient to know reaction rates and equilibrium constants: understanding the \emph{transport properties} of the biomolecules involved is also crucial \cite{hill}. For example, it is now known that many enzymes undergo enhanced diffusion as well as chemotaxis in the presence of their chemical substrates \cite{agud18c,jee17,dey14,agud18a,ried15,illi17a,jee18,xu19}. In turn, chemotaxis in response to chemicals that are being produced or consumed may lead to spontaneous self-organization of catalytic particles into chemically-active clusters \cite{agud19,wu15,zhao18}. Other works have shown the importance of segregation of different biomolecular components into phase-separated fluid compartments within the cell \cite{Boeynaems2018,schu18}, or how differences in diffusion coefficients between membrane-bound and cytosolic molecules are crucial for pattern formation and polarization in cells \cite{hala18b,hala18,brau18,wigb20,ramm18,gros18}.

One particularly ubiquitous feature of functional units in biology, be it proteins, enzymes, or molecular machines, is that they are oligomeric, i.e.~complexes composed of several subunits that can reversibly associate and dissociate \cite{trau94,dale99,ali05,andr08,levy08,coqu13,ahne15,garc17,gunt18,jee19,hida19}. These proteins are typically fully functional only in their oligomeric state.  One may thus wonder why oligomers are so prevalent, rather than highly stable proteins and protein complexes with irreversibly bound components. We note that, physically, reversibility implies that the associated binding energies and energy barriers are of the order of the thermal energy $k_\mathrm{B} T$.  Could there be, perhaps, a functional advantage to proteins being able to disassemble and reassemble?

\begin{figure*}[t]
	\centering
	\includegraphics[width=1\linewidth]{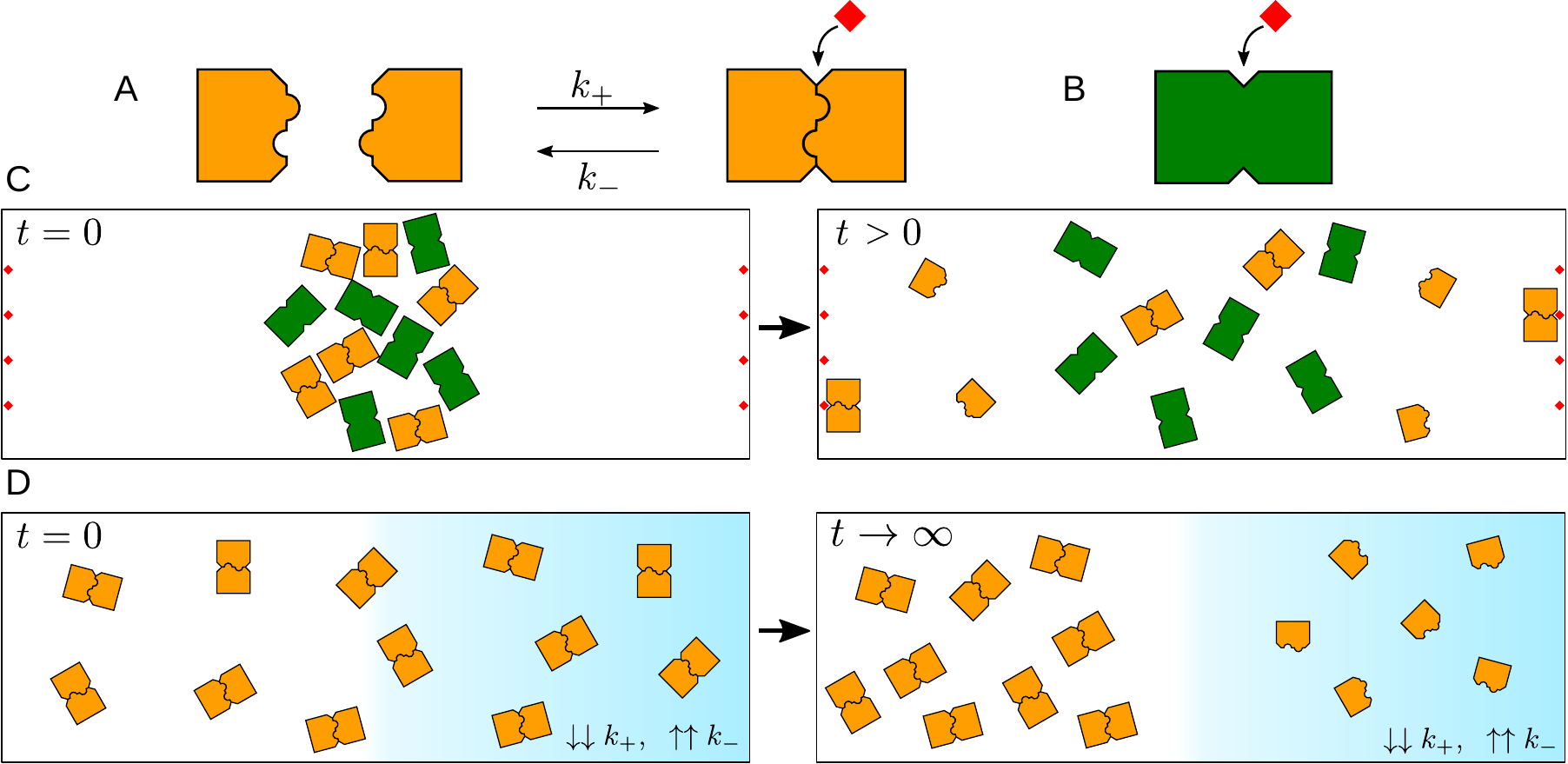}%or 17.8cm
	\caption{(A) Minimal model of an oligomeric protein. The monomers of a homodimeric protein can associate and dissociate with rates $k_+$ and $k_-$, which may be dependent on environmental conditions (concentration of salt or a chemical inhibitor, pH, illumination...). The protein is functional (in this case able to bind and react with the red ligand) only in its dimer form. (B) A non-dissociating but otherwise identical protein. (C) Faster diffusion of the monomers coupled to reassociation into dimers helps a dissociating protein reach a reactive target in shorter time than its non-dissociating counterpart. (D) In the presence of externally imposed spatial gradients of the dissociation rates, dissociating proteins undergo `stabilitaxis', i.e.~they tend to accumulate in regions where the oligomeric form is most stable. \label{fig_scheme}}
\end{figure*}

Inspired by this puzzle, we investigate here the transport properties of dissociating proteins (Fig.~\ref{fig_scheme}). One important question is how association-dissociation might affect the reactivity of a protein that needs to reach and react with a given target. Such problems, in which a protein diffuses until it finds a certain target, are typically known as \emph{first passage} problems, and have been subject of many studies in recent years. The effects of different spatial geometries and heterogeneous media \cite{cond07,beni10,matt12,guer16,gode16}, anomalous diffusion \cite{gitt00,rang00}, or intermittently switching transport kinetics of the protein \cite{love08,beni11,gode17,greb19}, on first passage times have all been explored to a certain extent. A common feature of all these studies, however, is that they deal with systems of non-interacting particles, in which each particle behaves independently from the others: the first passage time is thus related only to the transport properties of a single particle, and is independent of particle concentrations.

This is not the case for dissociating proteins; see Fig.~\ref{fig_scheme}\emph{C}. Indeed, whereas dissociation does occur independently for each protein, reassociation requires that two protein subunits find each other, and is thus dependent on the overall protein concentration in the system. The first passage time, therefore, becomes a \emph{collective} property of the system. In fact, we find that association-dissociation can lead to an enhancement in reactivity with respect to a stable non-dissociating protein, but this occurs \emph{cooperatively}, only for protein concentrations above a critical value. Enhanced reactivity due to association-dissociation is thus a markedly different phenomenon to that obtained in switching diffusion models \cite{gode17,greb19}, which represent e.g.~a protein undergoing conformational changes.

%\begin{figure*}[t]
%	\centering
%	\includegraphics[width=1\linewidth]{fig_scheme}
%	\caption{(A) Minimal model of an oligomeric protein. The monomers of a homodimeric protein can associate and dissociate with rates $k_+$ and $k_-$, which may be dependent on environmental conditions (concentration of salt or a chemical inhibitor, pH, illumination...). The protein is functional (in this case able to bind and react with the red ligand) only in its dimer form. (B) A non-dissociating but otherwise identical protein. (C) Faster diffusion of the monomers coupled to reassociation into dimers helps a dissociating protein reach a reactive target in shorter time than its non-dissociating counterpart. (D) In the presence of externally imposed spatial gradients of the dissociation rates, dissociating proteins undergo `stabilitaxis', i.e.~they tend to accumulate in regions where the oligomeric form is most stable. \label{fig_scheme}}
%\end{figure*}

A second important question with regards to the transport properties of oligomeric proteins is how they respond to heterogeneous environments, see Fig.~\ref{fig_scheme}\emph{D}. We demonstrate here that dissociating proteins tend to spontaneously accumulate in regions in which they are most stable, \emph{via} a generic mechanism which we term `stabilitaxis'. This behaviour may be exploited in order to trigger non-uniform patterns of protein in response to gradients of any stimuli that affects protein stability, be it concentration of a chemical inhibitor, salt, pH, or light.

The paper is organized as follows. In the Results section, we first describe the basic model for a dissociating homodimer protein, and predict enhanced diffusion and stabilitaxis arising from dissociation. We then show how enhanced diffusion coupled to reassociation leads to enhanced reactivity with localized targets through a cooperative mechanism, and demonstrate how stabilitaxis leads to non-uniform steady state patterns of protein in the presence of dissociation gradients. Finally, in the Discussion, we embed our results within the context of biology and materials engineering.

\section{Results}

\subsection{Enhanced diffusion and dissociation-induced drift velocity}

We consider the simplest model for the reversible association and dissociation of two identical monomers to form a homodimeric protein, see Fig.~\ref{fig_scheme}\emph{A}. The concentrations of monomer and of dimer, respectively $\ro$ and $\rt$, are governed by the coupled time evolution equations
\begin{eqnarray}
\partial_t \ro & = & D_1 \nabla^2 \ro - 2 \kpeff \ro^2 + 2 \kmeff \rt, \label{ro3} \\
\partial_t \rt  & = & D_2 \nabla^2 \rt + \kpeff \ro^2 - \kmeff \rt, \nonumber
\end{eqnarray}
where both the association and dissociation rates $\kpeff$ and $\kmeff$ can depend arbitrarily on the environmental conditions (e.g. concentration of salt or a chemical inhibitor, pH, illumination, etc.), which in turn may be space-dependent. The monomer diffuses with coefficient $D_1$, and the dimer with coefficient $D_2$. Note that, in general, the bulkier dimer will diffuse more slowly than the monomer, so that $D_2<D_1$. In fact, we have shown in previous work that, for two subunits that are linked into a dimer, the diffusion coefficient of the dimer goes as $D_2 = D_1/2 - \delta D_\mathrm{fluc}$ where $\delta D_\mathrm{fluc}>0$ corresponds to a fluctuation-induced hydrodynamic correction \cite{illi17b,adel19a,agud19b}. We therefore generically expect the even stronger condition $D_2<D_1/2$.

Direct analytical solution of the coupled non-linear evolution equations in Eq.~\textbf{\ref{ro3}} is hard. However, further progress can be achieved if we focus on the \emph{total} protein concentration $\re \equiv \ro/2 + \rt$, defined as the equivalent amount of dimeric protein, where the factor $1/2$ reflects the fact that two monomers are needed to generate a dimer. Summing both equations, we can write an evolution equation for the total protein concentration given by
\begin{equation}
\partial_t \re = \frac{D_1}{2} \nabla^2 \ro + D_2 \nabla^2 \rt.
\label{re}
\end{equation}
For sufficiently weak protein gradients, the typical timescale for diffusion is much slower than the association-dissociation timescale, and we can make a \emph{local equilibrium} approximation $\kpeff \ro^2 \approx \kmeff \rt$, implying that $\ro$ and $\rt$ quickly equilibrate at every point in space. Under this approximation, the local monomer and dimer concentrations are related to the local total protein concentration by
\begin{equation}
\ro \approx \frac{\Kd}{4}\left(\sqrt{1+16\frac{\re}{\Kd}}-1\right),~~~\text{and}~~~\rt \approx \frac{\ro^2}{\Kd},
\label{monoanddi}
\end{equation}
where we have defined the dissociation constant $\Kd \equiv \kmeff/\kpeff$, which carries the environment-dependence (or position-dependence) of the association and dissociation rates.

Inserting  the values resulting from the local equilibrium approximation into Eq.~\textbf{\ref{re}}, we  finally obtain an explicit evolution equation for the total protein concentration
\begin{equation}
\partial_t \re = \nabla \cdot \left( \Deff \nabla \re - \re \Vdis \right),
\label{re2}
\end{equation}
with the effective diffusion coefficient
\begin{equation}
\Deff \equiv D_2 + \frac{D_1-D_2}{\sqrt{1+16\re/\Kd}},
\label{Deff}
\end{equation}
and the dissociation-induced drift velocity
\begin{equation}
\Vdis \equiv - \frac{D_1-D_2}{8} \left( \frac{1+8\re/\Kd}{\sqrt{1+16\re/\Kd}}-1 \right) \frac{\nabla \Kd}{\re}.
\label{Vdis}
\end{equation}

Because the dimer diffuses more slowly than the monomer, with $D_2<D_1$, the effective diffusion coefficient is always larger than the dimer diffusion coefficient, $\Deff>D_2$, i.e.~dissociation leads to enhanced diffusion. The effective diffusion coefficient decreases monotonically with increasing protein concentration, from $\Deff=D_1$ at low protein concentration ($\re \ll \Kd$, in which case all proteins are in the form of monomers) to $\Deff=D_2$ at high protein concentrations ($\re \gg \Kd$, in which case all proteins are in the form of dimers). Equivalently, the effective diffusion coefficient increases monotonically with increasing $\Kd$ from $\Deff=D_2$ to $\Deff=D_1$.

Noting that the  coefficient multiplying $\nabla \Kd$ in Eq.~\textbf{\ref{Vdis}} is always negative, we see that the dissociation-induced velocity $\Vdis$ always points in the direction of decreasing $\Kd$, which is, towards regions where the dimer is more stable. We term this behavior \emph{stabilitaxis}. Moreover, we note that the magnitude of the velocity depends non-monotonically on the protein concentration, tending to zero for low ($\re \ll \Kd$) and high ($\re \gg \Kd$) protein concentrations, and reaching a maximum value at $\re \simeq 0.3 \Kd$.

	The approach that we just followed, based on using the local equilibrium approximation in order to obtain a closed evolution equation for the total protein concentration in a reaction-diffusion system with mass conservation, has been developed in great detail in the context of pattern formation by systems that exhibit a Turing instability \cite{hala18,brau18,wigb20}. In these studies, the reaction rates are typically position-independent (or at most, varying step-wise \cite{wigb20}) but the system itself can become laterally unstable. In the problem that we consider here, on the other hand, the association-dissociation rates may have an arbitrary space dependence, but the system would otherwise be laterally stable. The local equilibrium approximation was also used in the latter context in Ref.~\citenum{agud18a}, which studied enzyme chemotaxis in response to arbitrary substrate gradients.

\subsection{Cooperatively enhanced reactivity}

We have shown that the effective diffusion of the total amount (in both monomeric and dimeric form) of a dissociating protein is faster than that of a non-dissociating protein, i.e.~we always have $D_\mathrm{eff} > D_2$. This conclusion was to be expected, given that the smaller monomers will diffuse faster than the bulkier dimers. A less obvious question, and one more relevant to biology as well as technological applications, is whether association-dissociation can help a protein reach \emph{and} react with a distant reactive target more rapidly. Note that, while dissociation helps in enhancing diffusion, it also hinders the reaction by rendering the protein non-functional, which suggests a non-trivial competition between these two effects.

To this end, we have investigated the first passage time of dimers placed at the center of a one-dimensional domain of length $L$, with absorbing boundary conditions for the dimers [$\rt(x=0)=\rt(x=L)=0$] and no-flux boundary conditions [$\ro'(x=0)=\ro'(x=L)=0$] for the monomers. This represents a system in which a target located at the boundaries reacts instantaneously with dimers (diffusion-limited reaction), but is insensitive to monomers. The results for the dissociating case, obtained from numerical solution of the coupled partial differential equations in Eq.~\textbf{\ref{ro3}} with \emph{position-independent} $\kpeff$ and $\kmeff$, are compared with those for the diffusion of a non-dissociating protein, governed simply by $\partial_t \rtnd = D_2 \nabla^2 \rtnd$ (see Methods).

\begin{figure}%[t]
	\centering
	\includegraphics[width=1\linewidth]{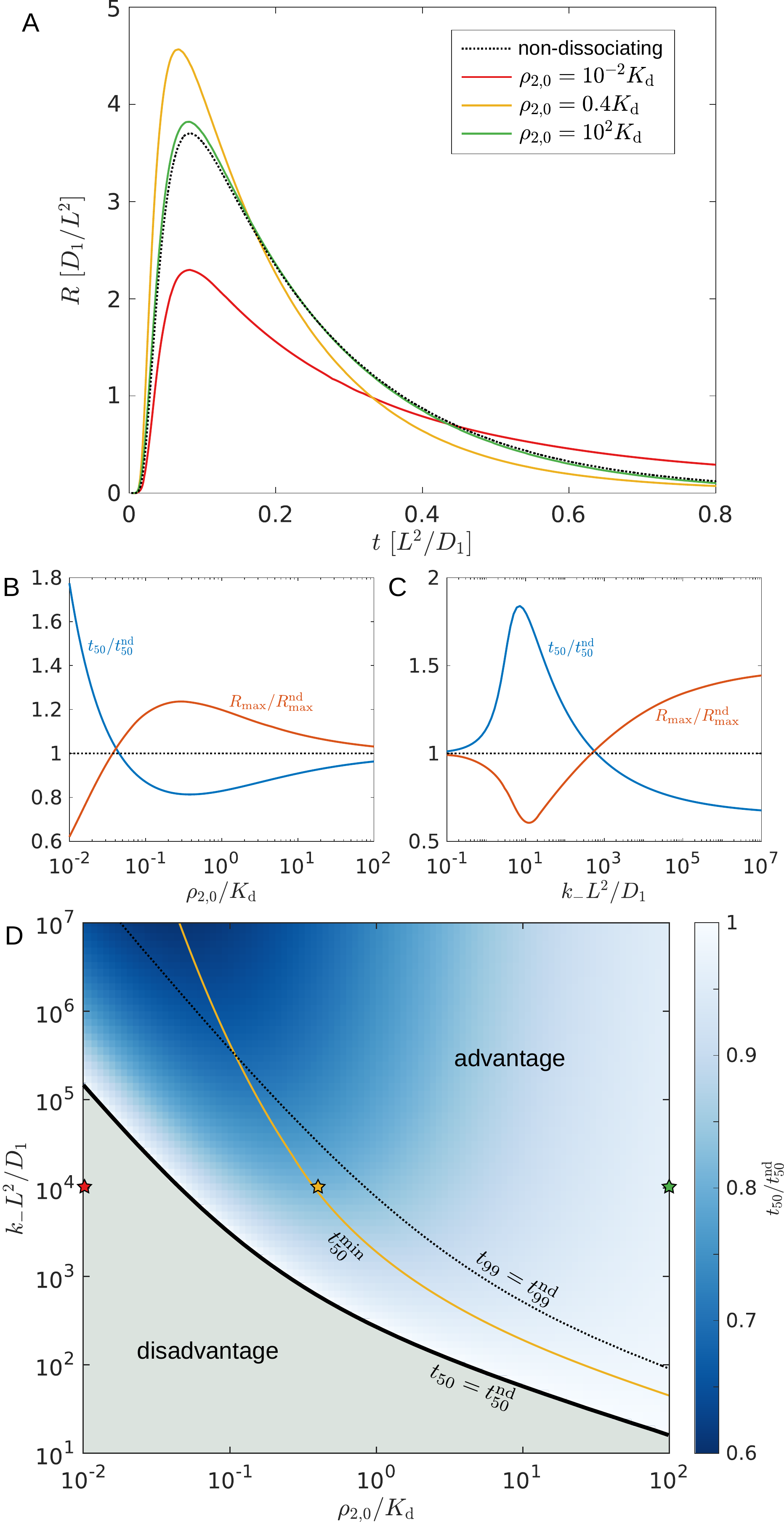}
	\caption{(A) Reaction rate as a function of time for a dissociating protein at three different concentrations, as well as that of a non-dissociating protein. (B and C) Median first passage time $t_{50}$ (time at which half of all proteins have reacted) and maximal reaction rate $R_\mathrm{max}$ relative to those of a non-dissociating protein $t_{50}^\mathrm{nd}$ and $R_\mathrm{max}^\mathrm{nd}$; (B) as a function of protein concentration; and (C) as a function of $k_- L^2/ D_1$ which compares the rate of association-dissociation to the diffusion rate. (D) $t_{50}/t_{50}^\mathrm{nd}$ as a function of both protein concentration and association-dissociation rate. The `disadvantage' region corresponds to $t_{50}/t_{50}^\mathrm{nd}>1$; the red, yellow, and green stars refer to the reaction rate curves in (A); the line labeled $t_{50}^\mathrm{min}$ indicates the concentrations that minimize $t_{50}$ for a given $k_-$; the line $t_{99}=t_{99}^\mathrm{nd}$ denotes the values above which $99\%$ of the proteins react faster in the dissociating case. In all cases we set $D_2=0.5D_1$; in (A and B) $k_- L^2/ D_1=10^4$; in (C) $\rho_{2,0}=0.4K_\mathrm{d}$.   \label{fig_advantage}}
\end{figure}

Because dimers are absorbed at the boundaries, the total protein number $N(t)=N_1(t)/2+N_2(t)$ within the box decreases with time. Here, $N_1$ and $N_2$ are the monomer and dimer numbers, with $N_i = \int_0^L \rho_i \mathrm{d}x$. We can then define a time-dependent \emph{reaction rate} as $R(t) = - \frac{1}{N(t=0)}\frac{\mathrm{d}N}{\mathrm{d}t}$. The reaction rate defined in this way verifies the normalization condition $\int_0^\infty R(t) \mathrm{d}t = 1$, and serves as a mean-field generalization of the first passage time probability distribution to a system with many interacting particles, which will coincide with the results of a stochastic approach in the limit of large number of particles.

We find that in a system of associating and dissociating particles, first passage is indeed a collective property of the system (Fig.~\ref{fig_advantage}). In particular, the reaction rate curve $R(t)$ depends on the total initial protein number, as given by the number $N_2(t=0) \equiv \rho_{2,0}L$ of dimers initially placed at the center of the box. The $R(t)$ curves for several values of $\rho_{2,0}$ are shown in Fig.~\ref{fig_advantage}\emph{A}, and compared with that of a non-dissociating protein (black dotted line). At low concentrations, the dissociating protein is mostly in monomer form, and reacts slower than a non-dissociating protein (red line). At intermediate values of protein concentration, however, a positive interplay between faster diffusion in the monomer state, coupled to sufficiently frequent reassociation into the reactive dimer state, leads to enhanced reactivity with respect to the non-dissociating protein (yellow line). As the protein concentration is further increased, the proteins spend most of the time in the dimer state and react with a very similar rate as a non-dissociating protein (green line).

Enhanced reactivity thus arises as a \emph{cooperative} effect from the interaction of a sufficiently large number of proteins. This is clearly seen in Fig.~\ref{fig_advantage}\emph{B}, which shows both the median first passage time $t_{50}$ and the peak reaction rate $R_\mathrm{max}$, relative to those of a non-dissociating protein $t_{50}^\mathrm{nd}$ and $R_\mathrm{max}^\mathrm{nd}$, as a function of protein concentration. The median first passage time obtained from $R(t)$ is a mean-field quantity representing the time after which $50\%$ of the initial proteins have reacted with the target, which, for a many-particle system such as the one under consideration, is a more intuitive measure of reaction speed than the mean first passage time. 

Moreover, we find that reactivity is enhanced when the dynamics of association-dissociation is sufficiently fast as compared to the diffusion timescale; see Fig.~\ref{fig_advantage}\emph{C}. For very slow dynamics, with $k_- L^2 / D_1 \ll 1$, the protein does not have time to dissociate before reaching the target, and thus behaves effectively as a non-dissociating protein. At intermediate values, dissociation is counterproductive, as the protein has sufficient time to dissociate before reaching the target, but still takes a long time to reassociate in order to react. Finally, when the dynamics becomes sufficiently fast, dissociation is always favorable as it enhances diffusion (Eq.~\textbf{\ref{Deff}}) while reassociation is fast enough to not hinder the reaction.

The combined effect of protein concentration and association-dissociation dynamics on the median first passage time is summarized in Fig.~\ref{fig_advantage}\emph{D}, for the particular case $D_2=0.5D_1$. Cooperatively enhanced reactivity is found at an intermediate range of protein concentrations and for sufficiently fast association-dissociation dynamics. The optimal value of concentration that minimizes the median first passage time decreases with increasing $k_-$ (yellow line). Within the range of values explored, the median first passage time can be up to 40\% smaller for a dissociating protein than for a non-dissociating protein, and will decrease even further for larger values of $k_- L^2 / D_1$. Note that our results remain qualitatively similar if a measure of reaction speed other than the median first passage time is used. As an example, we also show the line $t_{99}=t_{99}^\mathrm{nd}$ (dotted line), representing the values above which the time it takes for $99\%$ of the proteins to react is shorter for a dissociating protein than for a non-dissociating one.

The enhancement in reactivity (reduction in median first passage time) that can be achieved due to dissociation increases as the ratio $D_2/D_1$ is decreased; see Fig.~S1 for the case $D_2/D_1=0.3$. In fact, we expect that the minimal median first passage time that can be achieved is $t_{50}=(D_2/D_1) t_{50}^\mathrm{nd}$, which will occur in the limit in which the protein concentration is very low $\rho_{2,0}\ll K_\mathrm{d}$. In this regime, the protein is mostly in monomer form. However, since the association and dissociation rates are very fast, namely $k_- L^2 / D_1 \gg 1$, reassociation can occur very rapidly near the target.  Note that all the results just described were obtained from numerical solution of the full evolution equations (Eq.~\textbf{\ref{ro3}}) at finite $k_- L^2 / D_1$. In order to examine the limit of very fast association-dissociation, we can instead consider the first passage time problem using the local equilibrium approximation in Eq.~\textbf{\ref{re2}}, which in fact corresponds to the limit $k_- L^2 / D_1 \to \infty$. We have solved this equation numerically, for the case $D_2=0.5D_1$, to obtain the median first passage time as a function of total protein concentration, and indeed we find that the first passage time goes from that expected of a monomer ($t_{50}=0.5t_{50}^\mathrm{nd}$) at very low concentration, to that expected of a dimer ($t_{50}=t_{50}^\mathrm{nd}$) at very high concentration (Fig.~S2).  Note that the limit $k_- L^2 / D_1 \to \infty$ corresponding to the local equilibrium approximation shows qualitatively different behavior to that seen at finite $k_- L^2 / D_1$, because in this limit there is no optimal concentration for which the median first passage time is minimal; instead, the first passage time increases monotonically with increasing concentration. This implies that the optimal concentration tends to zero as $k_- L^2 / D_1$ tends to infinity.

	The results just described were obtained for the first passage time of dimers initially placed at the center of a one-dimensional domain with reactive boundaries, but our results hold more generally. In particular, we find that dimensionality does not play a role, and dimers placed at the center of a two-dimensional circular domain or a three-dimensional spherical domain with reactive boundaries also show cooperatively enhanced reactivity, with nearly identical enhancements; see Fig.~S3. Moreover, cooperatively enhanced reactivity is also robust to the choice of initial conditions for the monomer and dimer distributions. As a particularly relevant example, we have considered, as initial condition, a laterally uniform distribution of monomers and dimers at association-dissociation equilibrium ($\rho_2 = \rho_1^2/\Kd$), instead of a highly-concentrated distribution of dimers at the center of the domain. This would correspond to a case in which the system is first allowed to relax to equilibrium in the absence of the boundary reaction, and then the boundary reaction is switched on. We find that there is also cooperatively enhanced reactivity for this choice of initial conditions, with very similar enhancements as above; see Fig.~S4.

\subsection{Stabilitaxis: accumulation in regions of higher stability}

The existence of the dissociation-induced drift velocity (Eq.~\textbf{\ref{Vdis}}) suggests that, in environments with \emph{position-dependent} dissociation, dissociating proteins will tend to preferentially accumulate in the regions of higher stability after some time. Indeed, we can verify the existence of such `stabilitaxis' by calculating the steady state concentrations for the monomer, dimer, and total amount of protein in a non-uniform environment. From Eq.~\textbf{\ref{re}}, we see that the total flux of protein is given by $\boldsymbol{J} = - \nabla ( D_1 \rho_1 / 2 + D_2 \rho_2 )$. Requiring that this flux be equal to zero, $\boldsymbol{J} = 0$, we find that in a steady state with no influx or outflux of proteins into the system, the combination $D_1 \rho_1 / 2 + D_2 \rho_2$ must be a position-independent constant. Combining this condition with the results of the local equilibrium approximation in Eq.~\textbf{\ref{monoanddi}}, we finally find the steady-state profiles
\begin{eqnarray}
\rho_{1,\infty} \approx  \frac{\Kd}{4} \frac{D_1}{D_2} \left( - 1 + \sqrt{ 1 + \frac{C}{\Kd}} \right), \nonumber  \\
\rho_{2,\infty}  \approx  \frac{\rho_{1,\infty}^2}{\Kd},~~\text{and}~~\rho_\mathrm{tot,\infty} \approx   \frac{\rho_{1,\infty}}{2} + \frac{\rho_{1,\infty}^2}{\Kd} ,
\label{steady}
\end{eqnarray}
where $C$ is a constant with units of concentration, which is used to satisfy the constraint on the total amount of protein.
The same approach, based on combining the local equilibrium approximation with the condition of zero total protein flux at steady state, has been recently used to understand pattern formation in reaction-diffusion systems that display a lateral instability, including a novel geometric interpretation of the concepts of local equilibrium and zero total flux, both in the case of spatially-uniform reaction rates \cite{brau18} and in environments with a step-wise position-dependence of the reaction rates \cite{wigb20}.

\begin{figure}%[t]
	\centering
	\includegraphics[width=1\linewidth]{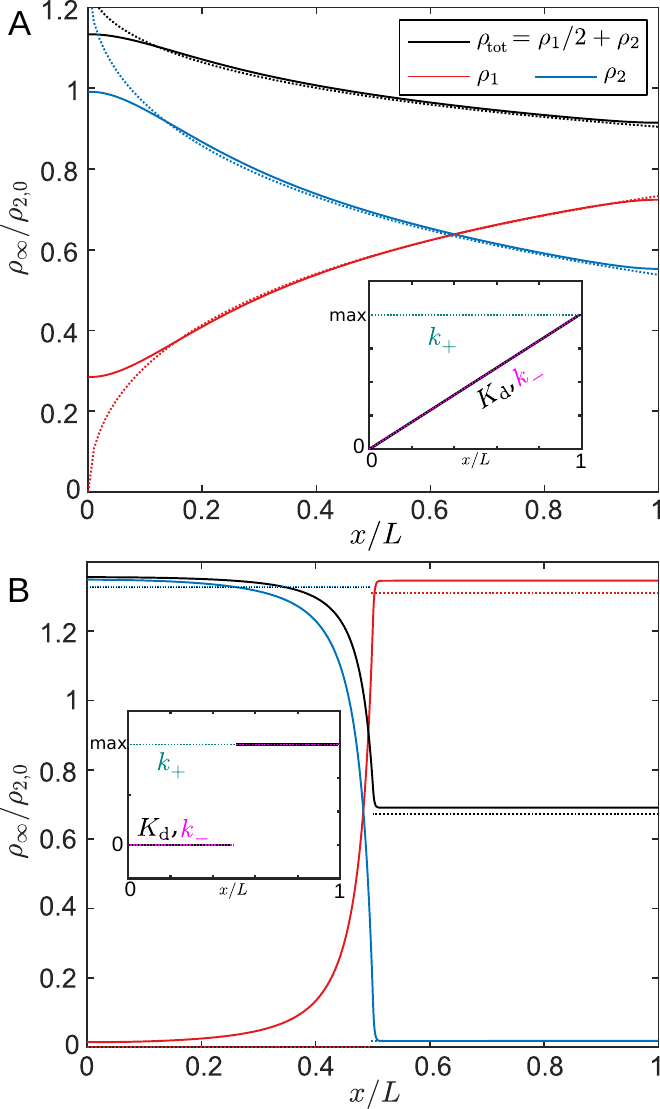}
	\caption{Steady-state concentrations for a protein in the presence of a dissociation gradient. The numerical solution of Eq.~\textbf{\ref{ro3}} (solid lines) can be compared to the local equilibrium approximation in Eq.~\textbf{\ref{steady}} (dotted lines). The protein undergoes stabilitaxis, accumulating in regions of higher stability. The insets show the corresponding dissociation gradients (arbitrary units): (A) linear gradient in $\kmeff$ leading to linearly increasing $\Kd$; and (B) discrete jump in $\kmeff$ and thus $\Kd$. In both cases we have set $D_2=0.5D_1$. In (A) $k_-^\mathrm{max} L^2 / D_1 = 10^2$ and $\rho_{2,0}=K_\mathrm{d}^\mathrm{max}$. In (B) $k_-^\mathrm{max} L^2 / D_1 = 10^5$ and $\rho_{2,0}=10^{-2} K_\mathrm{d}^\mathrm{max}$. \label{fig_grads}}
\end{figure}

To confirm the validity of our steady-state results, we have compared them to the long time limit of the numerical solution of the coupled partial differential equations (Eq.~\textbf{\ref{ro3}}), with no-flux boundary conditions $\ro'(x=0)=\rt'(x=0)=\ro'(x=L)=\rt'(x=L)=0$ for all species, for two different examples of position-dependent association and dissociation rates $\kpeff$ and $\kmeff$, which naturally lead to a position-dependent $\Kd = \kmeff/\kpeff$ (Fig.~\ref{fig_grads}). The steady state profile given by Eq.~\textbf{\ref{steady}} reproduces well the numerical results, although it deviates near the box boundaries. This can be understood by noting that the no-flux boundary conditions are not appropriately captured by the local equilibrium approximation, as well as regions with sharp changes in $K_\mathrm{d}$  and thus in the protein concentration. The width of the region over which deviations are significant is governed by the length scale $\sqrt{D_1/k_-}$, as can be understood from a first order approximation valid for small deviations around the local equilibrium approximation (see the Supplementary Information). Therefore, these deviations become progressively smaller, and the local equilibrium approximation increasingly more accurate, with increasing $k_- L^2 / D_1$ (see Fig.~S5). As predicted, the protein does preferentially accumulate in regions of higher stability (lower $\Kd$), whether one considers the total protein amount including monomer and dimer forms (black lines), or just the dimer form (blue lines). These results are independent of the dimensionality of the system, and we obtain similar profiles, well captured by the local equilibrium approximation, for two- and three-dimensional circular and spherical domains (Fig.~S6). The fact that the steady-state profiles depend on the ratio of diffusion coefficients (see Eq.~\textbf{\ref{steady}}) clearly demonstrates that stabilitaxis is a non-equilibrium phenomenon, which must be sustained by externally imposed gradients.

The dependence of stabilitaxis on the ratio of diffusion coefficients is most evident if we consider the limit of a space containing two connected regions; one with very weak dissociation $K_\mathrm{d} \to 0$, for which we expect $\rho_\mathrm{tot} \approx \rt$ and $\ro\approx0$, and the other with very strong dissociation $K_\mathrm{d} \to \infty$, for which we expect $\rho_\mathrm{tot} \approx \ro/2$ and $\rt \approx 0$. Taking these limits in Eq.~\textbf{\ref{steady}}, and solving for $C$, we obtain the relation
\begin{equation}
D_2 \rho_\mathrm{tot,\infty}(K_\mathrm{d} \to 0) \approx D_1 \rho_\mathrm{tot,\infty}(K_\mathrm{d} \to \infty),
\label{steady2} 
\end{equation}
between the protein concentrations in both regions. For $D_2=0.5D_1$, we thus expect twice as much protein in the region where it is stable as compared to the region where it is unstable. This prediction is confirmed by the numerical result in Fig.~\ref{fig_grads}\emph{B}.

	In order for a protein gradient to be established \emph{via} stabilitaxis, the underlying dissociation gradient should be sufficiently long lived. The typical timescale over which the steady state distribution is reached is given by $L^2/D_1$, and thus the required minimal lifetime of the underlying gradient will strongly depend on the system size. Using $D_1 = 10$ $\mu$m$^2$/s as a typical protein diffusion coefficient in the cytoplasm, we find characteristic times of $0.1$~s for a small cell with $L=1$~$\mu$m, $10$~s for a cell with $L=10$~$\mu$m, or $1000$~s for a large cell with $L=100$~$\mu$m.

\section{Discussion}

We have predicted here a number of non-trivial features in the diffusion, reactivity, and gradient response of dissociating oligomeric proteins. Some of these features could be seen as conferring a functional advantage to dissociating proteins over non-dissociating ones, which might explain why biological evolution has resulted in many important enzymes and proteins being multimeric.

Firstly, we have shown that it can be advantageous for a protein, enzyme, or molecular machine to dissociate into non-functional but smaller subunits that can diffuse faster, and later reassociate to perform their function at a distant location. This can lead to significantly faster reaction rates for dissociating proteins. We have shown that enhanced reactivity arises as a cooperative effect, which minimizes reaction time for an intermediate range of protein concentrations.  Moreover, as can be seen from Fig.~\ref{fig_advantage}, dissociation becomes more and more advantageous with increasing values of the dimensionless quantity $k_- L^2 / D_1$, which compares the rate of unbinding with the typical timescale of diffusion across the system. Crucially, this quantity scales with the square of the system size, and therefore can vary over many orders of magnitude for different systems.  Experimental measurements of protein complex dissociation kinetics are not widely available, but some relevant examples can be found, such as $k_- \approx 20$~s$^{-1}$ for the CheY-CheA complex involved in the sensory response pathway of \emph{E.~coli} \cite{stew04}, or $k_- \approx 2$~s$^{-1}$ for the p53-Mdm2 complex involved in DNA repair \cite{scho02} as well as for the WASp-Cdc42 complex involved in the remodelling of actin filaments \cite{hems05}. Using a moderate choice of $k_-=1$~s$^{-1}$ for the dissociation rate \cite{schr02,schr09,kuma09},  and $D_1=10$~$\mu$m$^2/$s for the diffusion coefficient of a protein in the cytoplasm, we find values of $k_- L^2 / D_1$ ranging from $10^{-1}$ for a small cell with $L=1$~$\mu$m, to $10^3$ for a large cell with $L=100$~$\mu$m, all the way up to $10^7$ for diffusion along a neuronal axon or a microfluidic device with $L=1$~cm. For membrane-bound proteins, the diffusion coefficient is greatly reduced, and thus the corresponding values of $k_- L^2 / D_1$ will be significantly enhanced, and the advantages due to dissociation will be greater. For any given protein, the advantages due to dissociation will be largest when the target to be reached is distant. The typical reactivity enhancements that can be achieved are of the order of $D_1/D_2$, and thus for a dissociating dimer are of the order of 10--50$\%$ (Fig.~\ref{fig_advantage}).

Secondly, we have shown that dissociation provides a new mechanism for proteins to sense and respond to their environment, by undergoing \emph{stabilitaxis} or motion towards regions in which their oligomeric form is most stable. Stabilitaxis represents a new way by which non-uniform patterns in the concentration of a biomolecule can be triggered. For example, polarization in the concentration of a dissociating protein within a cell can be triggered by localized production of a chemical that enhanced or inhibits the association or dissociation of the protein subunits. The precise form of the resulting protein distribution can be predicted from Eq.~\textbf{\ref{steady}}, but in general the typical difference in protein concentration between the regions of low and high dissociation will be proportional to $D_1/D_2$; see Eq.~\textbf{\ref{steady2}}. It remains to be seen whether stabilitaxis is exploited by the cell in the intracellular organization of oligomeric proteins. 

Although we have focused here for simplicity on the case of a homodimeric protein, we expect that our general predictions of enhanced reactivity and stabilitaxis will hold equally for more complex cases of hetero-multimeric proteins (i.e.~composed of more than two subunits, that may also be different from each other). As an example of a more complex protein, we have considered a homohexamer, composed of six identical subunits, and found qualitatively similar results both for enhanced reactivity and stabilitaxis (Supplementary Information, Fig.~S7). Interestingly, our numerical results show that the prediction in Eq.~\textbf{\ref{steady2}} for stabilitaxis still holds, if we replace the dimer diffusion coefficient by the hexamer diffusion coefficient. Because the ratio of monomer and hexamer diffusion coefficients is much larger (of the order of 6) the protein accumulation due to stabilitaxis is enhanced significantly. We expect that the maximum achievable enhancement in reactivity (for sufficiently fast association-dissociation rate) will also be larger for a multimeric protein with a large difference in diffusion coefficient between the monomeric and multimeric forms.

Beyond the biological implications, our predictions of enhanced reactivity may be useful in the context of chemical engineering, e.g.~in the design of synthetic catalytic microreactors. Moreover, our results may also be tested and applied in purely synthetic systems, e.g.~using patchy colloids coated with ligands, that can bind to each other to form colloidal molecules. In the context of engineering of active or responsive materials, one particularly interesting application would be to use colloids coated with light-induced linkers \cite{gunt15} that bind to each other only when illuminated. Such a material would flow and become denser in illuminated regions on demand.

\textbf{Acknowledgements.} We thank Jean-Fran\c cois Rupprecht for comments on an early version of the manuscript. J.A-C. and R.G. acknowledge funding from the U.S. National Science Foundation under MRSEC Grant No. DMR-1420620.

\section{Methods}

\textbf{Numerical solution of evolution equations.} The coupled evolution equations (\ref{ro3}) are numerically solved using MATLAB's \emph{pdepe} solver for systems of parabolic partial differential equations \cite{skee90}.  The size of the system is given by $L$, which in 1-D calculations corresponds to the length of the domain, and in 2-D and 3-D calculations to the the diameter of the circular or spherical domain. We can then define the dimensionless time as $\tau \equiv tD_1/L^2$, position as $\tilde{x}\equiv x/L$ in 1-D or as the radial coordinate $\tilde{r} \equiv r/L$ in 2-D and 3-D, and concentrations as $\tilde{\rho} \equiv (\kpeff/\kmeff)\rho = \rho/K_\mathrm{d}$. The system is then governed by two dimensionless parameters only, namely the ratio of association-diffusion timescales $\tkmeff \equiv \kmeff L^2 / D_1$, and the ratio of dimer-to-monomer diffusion coefficients $\tD \equiv D_2/D_1$, as well as our choice of initial conditions. In cases with position-dependent $k_-$, we use the maximum value $k_-^\mathrm{max}$ for the non-dimensionalization. For the initial conditions, we use (except for Fig.~S4) a Gaussian profile located at the center of the box for the concentration of the dimer, with standard deviation $\sigma=0.01L$, and normalized so that the total amount of dimer in the box is $\rho_{2,0} L$; the initial concentration of the monomers is set to zero. We use 1000 points in the space discretization. The system is evolved in time until $99\%$ of the proteins have been consumed (when calculating the reaction rate), or until a steady state is reached (when exploring stabilitaxis).

\bibliography{biblio}

%\clearpage

\clearpage
\onecolumngrid
\appendix

\section{Supplementary Information for \\ ``Cooperatively enhanced reactivity and `stabilitaxis' of dissociating oligomeric proteins"}

\renewcommand{\theequation}{S\arabic{equation}}
\setcounter{equation}{0}

\subsection{Calculations for a dissociating hexamer}

We consider a model of a dissociating hexamer, composed of monomers which can disassemble and reassemble into all of the intermediate states of dimer, trimer, tetramer, and pentamer. All reactions $[m] + [n]\xrightleftharpoons[\km]{\kp}[m+n]$  for association of a $m$-mer and a $n$-mer into a $(m+n)$-mer, as well as the corresponding reverse dissociation reactions, are considered as long as $m+n \leq 6$. For simplicity, all reactions are taken to occur with the same association and dissociation rates $k_+$ and $k_-$, which thus gives the same dissociation constant $K_\mathrm{d}\equiv \km/\kp$ for all reactions. The concentration of $m$-mers is denoted as $\rho_m$, and their diffusion coefficient by $D_m$. The system is then described by the six coupled reaction-diffusion equations
\begin{eqnarray}
\partial_t \ro & = & D_1 \nabla^2 \ro + \km (2\rt+\rho_3+\rho_4+\rho_5+\rho_6) - \kp \rho_1 (2\rho_1+\rho_2+\rho_3+\rho_4+\rho_5), \nonumber \\
\partial_t \rt & = & D_2 \nabla^2 \rt + \km(\rho_3+2\rho_4+\rho_5+\rho_6)+\kp\ro^2-\km\rt-\kp\rt(\ro+2\rt+\rho_3+\rho_4), \nonumber \\
\partial_t \rho_3 & = & D_3 \nabla^2 \rho_3 + \km(\rho_4+\rho_5+2\rho_6)+\kp\ro\rt-\km\rho_3-\kp\rho_3(\rho_1+\rho_2+2\rho_3), \nonumber \\
\partial_t \rho_4 & = & D_4 \nabla^2 \rho_4 + \km(\rho_5+\rho_6) +\kp(\ro\rho_3+\rt^2) -2\km  \rho_4-\kp\rho_4(\ro+\rt), \nonumber \\
\partial_t \rho_5 & = & D_5 \nabla^2 \rho_5 + \km\rho_6+\kp(\ro\rho_4+\rt\rho_3)-2\km  \rho_5-\kp\ro\rho_5, \nonumber \\
\partial_t \rho_6 & = & D_6 \nabla^2 \rho_6  + \kp(\ro\rho_5+\rt\rho_4+\rho_3^2)-3\km\rho_6, \nonumber
\end{eqnarray}
which satisfy a local conservation law for the total protein concentration (defined as the equivalent hexamer concentration)
\begin{equation}
\rho_\mathrm{tot} \equiv \frac{1}{6}(\ro+2\rt+3\rho_3+4\rho_4+5\rho_5+6\rho_6), \nonumber
\end{equation}

For the numerical calculation of the first passage time of the hexamer, we set no-flux boundary conditions for the monomer, dimer, trimer, tetramer, and pentamer, and absorbing boundary conditions for the hexamer. For the numerical calculation demonstrating stabilitaxis, we set no-flux boundary conditions for all species. In all cases, the diffusion coefficients are set as $D_n = D_1/n$, e.g.~the hexamer diffuses six times more slowly than the monomer. The results are shown in Fig.~S3.

\subsection{Boundary corrections in steady-state profiles}
As mentioned in the main text, and seen on Fig.~3, the local equilibrium approximation is rather good except at the boundaries of the domain (in the case of a linear dissociation profile) or at the location of the step (in the case of a step-wise dissociation profile). We find that these deviations become progressively smaller with increasing $k_- L^2/D_1$. In particular, the width of the region in which the deviations from the local equilibrium approximation occur appears to decrease with increasing $k_- L^2/D_1$; see Fig.~S5. This suggests that the width of these regions is controlled by the length scale $\sqrt{D_1/k_-}$. The regions of deviation from the local equilibrium approximation are expected to arise because this approximation cannot capture the individual no-flux boundary conditions for the monomer and dimer species at the boundaries, or the continuity of the protein concentrations across the dissociation step.

As described in the main text, the condition for zero total flux of protein in the steady state is $D_1 \rho_1 + 2D_2 \rho_2 = \tilde{C}$, where $\tilde{C}$ is a constant. Solving for $\rho_2$ and substituting this value into Eq.~\textbf{1} in the main text for the monomer distribution, and looking for the steady state with $\partial_t \rho_1 = 0$, gives an exact equation for the monomer distribution at steady state which reads
\begin{equation}
D_1 \rho_1'' - 2k_+ \rho_1^2 + \frac{k_-}{D_2} (\tilde{C} - D_1 \rho_1) = 0,
\label{eq1}
\end{equation}
where we focus on the 1-D case for simplicity, and the prime ($'$) denotes a derivative with respect to position. The local equilibrium approximation, which we will denote as $\bar{\rho}_1$, corresponds to the solution of the reactive part
\begin{equation}
- 2k_+ \bar{\rho}_1^2 + \frac{k_-}{D_2} (\tilde{C} - D_1 \bar{\rho}_1) = 0,
\label{eq2}
\end{equation}
which results in Eq.~[7] in the main text. In order to gain insight into the deviations from the local equilibrium approximation, we now consider the deviations $\delta \rho_1$ defined by $\rho_1 = \bar{\rho}_1 + \delta \rho_1$. Introducing this into Eq.~\ref{eq1} above, and using Eq.~\ref{eq2} above, we obtain
\begin{equation}
\delta \rho_1''= \frac{k_-}{D_1} \left( 2 \frac{\delta \rho_1^2}{\Kd} + 4 \frac{\bar{\rho}_1\delta \rho_1}{\Kd} + \frac{D_1}{D_2} \delta \rho_1 \right) - \bar{\rho}_1''.
\label{eq3}
\end{equation}
For small deviations from the local equilibrium approximation, the term quadratic in $\delta \rho_1$ can be neglected, and moreover, for large $k_- L^2/D_1$, the last term can also be neglected, so that we obtain the first order approximation
\begin{equation}
\delta \rho_1'' \approx \frac{k_-}{D_1} \left( 4 \frac{\bar{\rho}_1}{\Kd} + \frac{D_1}{D_2} \right) \delta \rho_1.
\label{eq4}
\end{equation}
Note that, in principle, $k_-$, $\bar{\rho}_1$, and $\Kd$ all depend on position. However, if their variation along the length scale $\sqrt{D_1/k_-}$ is small, we can assume them to be locally constant, and obtain the simple solution around any given location $x=x_0$
\begin{equation}
\delta \rho_1 \approx c_1 e^{(x-x_0)/\ell(x_0)} + c_1 e^{-(x-x_0)/\ell(x_0)},
\label{eq5}
\end{equation}
where $c_1$ and $c_2$ are constants to be determined by the boundary conditions, and $\ell(x_0)$ is a decay length given by
\begin{equation}
\ell(x_0) \equiv \left[ \frac{k_-(x_0)}{D_1} \left( 4 \frac{\bar{\rho}_1(x_0)}{\Kd(x_0)} + \frac{D_1}{D_2} \right) \right]^{-\frac{1}{2}},
\label{eq6}
\end{equation}
where $k_-$, and $\bar{\rho}_1/\Kd$ are evaluated at $x=x_0$.

We can now use this solution to obtain the corrections at the domain boundaries or at the step in the case of step-wise dissociation. In the case of domain boundaries, we need to satisfy the no-flux boundary condition $\rho_1'(x=x_b)=0$, where $x_b=0$ or $x_b=L$ corresponds to the left or right boundary, respectively. This condition in turn implies $\delta \rho_1'(x=x_b) = - \bar{\rho}_1'(x=x_b)$. Enforcing this condition and the fact that the correction should decay away from the boundary, we finally obtain the corrections accounting for the no-flux condition at the left and right boundaries which can be summed up to obtain the full correction
\begin{equation}
\delta \rho_1 \approx \ell(0) \, \bar{\rho}_1'(0) \, e^{-x/\ell(0)} - \ell(L) \, \bar{\rho}_1'(L) \, e^{(x-L)/\ell(L)}.
\label{eq7}
\end{equation}
In the case of a discontinuous change in dissociation at the location of the step $x=x_s$, the boundary condition that we must enforce is the continuity of the protein concentration at the step, i.e.~the values coming from the left and from the right must coincide $\rho_1(x \to x_s^-) = \rho_1(x \to x_s^+)$. Enforcing this boundary condition, and the fact that the correction should decay away from the step, we obtain the correction
\begin{equation}
\delta \rho_1 \approx \frac{\bar{\rho}_1^{(R)}-\bar{\rho}_1^{(L)}}{1+\ell^{(L)}/\ell^{(R)}} \left( \frac{\ell^{(L)}}{\ell^{(R)}} e^{(x-x_s)/\ell^{(L)}} [1 - \Theta(x-x_s)] - e^{-(x-x_s)/\ell^{(R)}}\Theta(x-x_s)  \right) ,
\label{eq8} 
\end{equation}
where the superscripts $(L)$ and $(R)$ indicate that the functions are evaluated at the left or at the right  of the step, respectively, and $\Theta(x)$ is the Heaviside step function. In both cases, the first order correction to the dimer distribution is obtained from the condition of zero total flux of protein in the steady state described above, which gives $\delta \rho_2 = -\frac{D_1}{2D_2} \delta \rho_1$.

This first order approximation is compared to the full numerical results, as well as the local equilibrium approximation, in Fig.~S5. As would be expected, the first order approximation performs better than the local equilibrium approximation. Moreover, the first order approximation rather accurately captures the decreasing width of the boundary deviations with increasing $k_- L^2/D_1$.

We note that, unfortunately, this simple analysis of the boundary deviations breaks down for the cases examined in Fig.~3 of the main text, as in this case the monomer concentration in the local equilibrium approximation $\bar{\rho}_1$ tends to zero at the left boundary (Fig.~3A) or at the left side of the step (Fig.~3B). As a consequence, the linearization given by Eq.~[4] is not valid, and we must work with the quadratic term in Eq.~[3]. The analysis thus becomes more involved and is beyond the scope of this paper. Nevertheless, the width of the boundary deviations is still controlled by the length scale $\sqrt{D_1/k_-}$, and thus decreases with increasing $k_- L^2/D_1$.

\clearpage

\renewcommand\thefigure{S\arabic{figure}}
\setcounter{figure}{0}

  \begin{figure}%[t]
 	\centering
 	\includegraphics[width=0.8\linewidth]{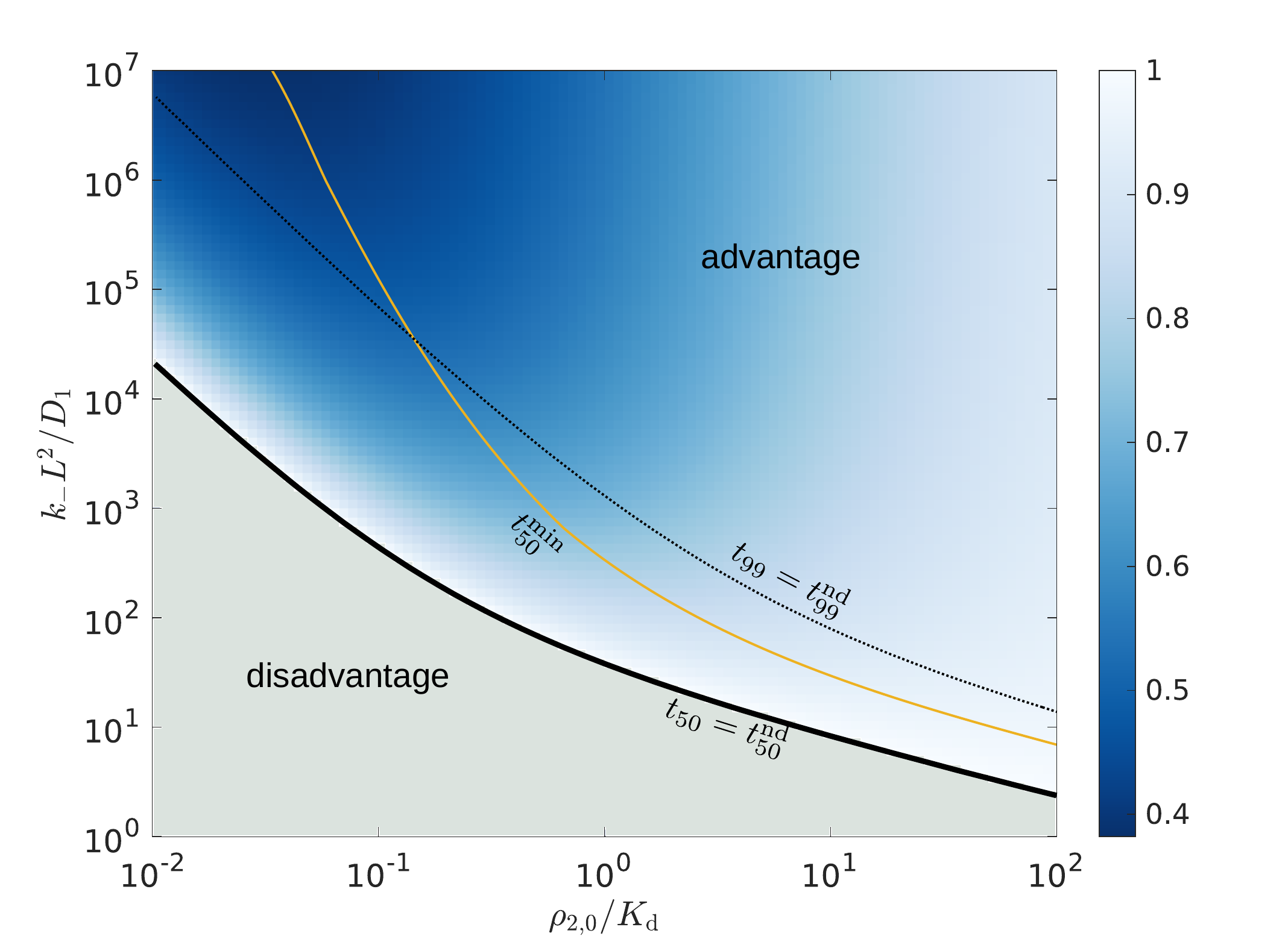}
 	\caption{Cooperatively enhanced reactivity of dissociating dimers with $D_2=0.3D_1$. $t_{50}/t_{50}^\mathrm{nd}$ as a function of both protein concentration and association-dissociation rate. The `disadvantage' region corresponds to $t_{50}/t_{50}^\mathrm{nd}>1$; the line labeled $t_{50}^\mathrm{min}$ indicates the concentrations that minimize $t_{50}$ for a given $k_-$; the line $t_{99}=t_{99}^\mathrm{nd}$ denotes the values above which $99\%$ of the proteins react faster in the dissociating case. Comparing to Fig.~2 in the main text, we see that the advantages due to dissociation are bigger when $D_2$ is decreased.  \label{fig_sm_0p3}}
 \end{figure}
 
 \begin{figure}%[t]
 	\centering
 	\includegraphics[width=0.7\linewidth]{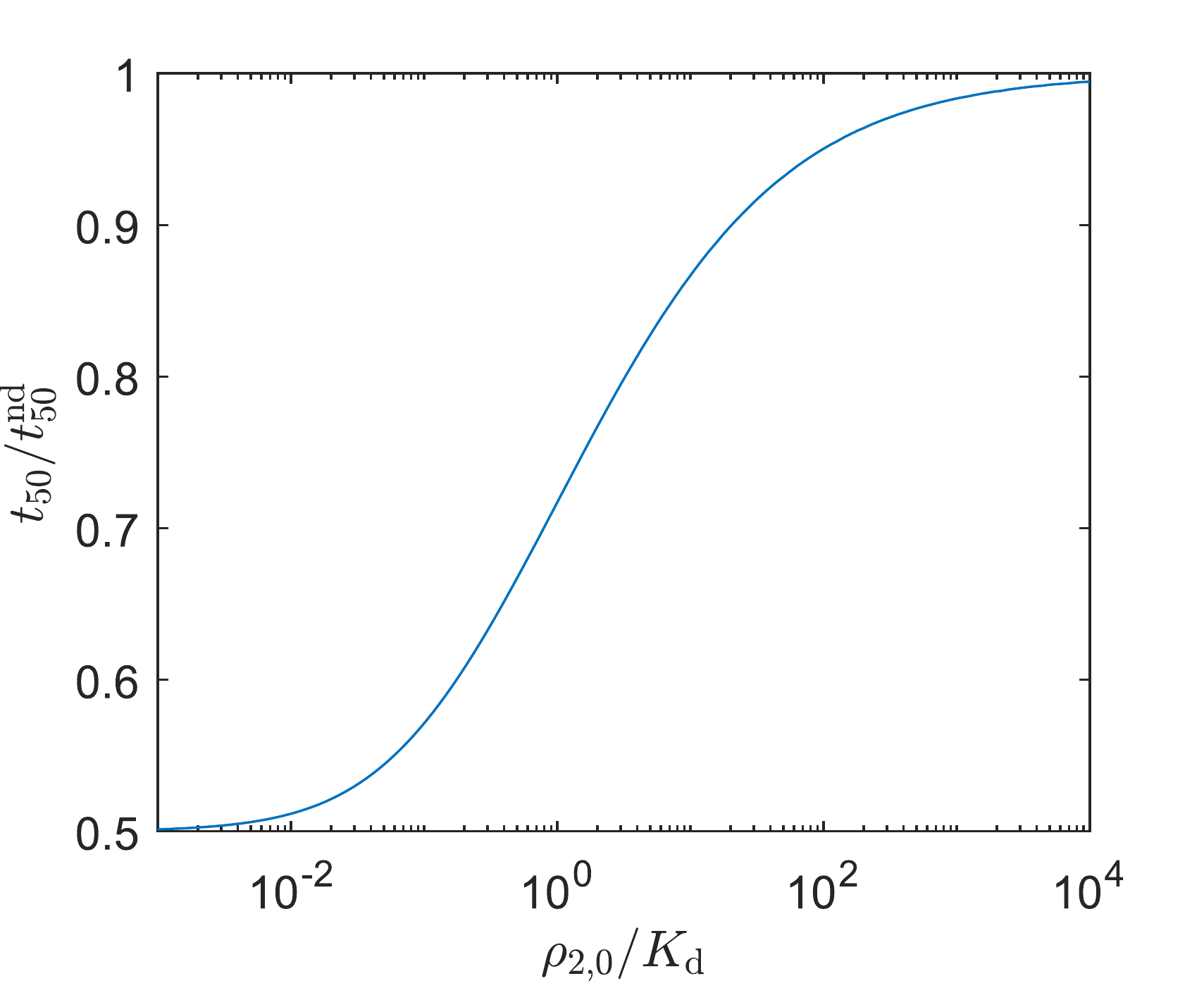}
 	\caption{Enhanced reactivity as a function of protein concentration in the limit of very fast association-dissociation. Median first passage time as obtained from solution of Eq.~\textbf{4} in the main text with absorbing boundary conditions $\rho_\mathrm{tot}(x=0)=\rho_\mathrm{tot}(x=L)=0$, which represents the limit $k_- L^2 /D_1 \to \infty$. We have used $D_2=0.5D_1$ and no dissociation gradient ($\nabla K_\mathrm{d}=0$). The median first passage time goes from that expected of a monomer ($t_{50}=t_{50}^\mathrm{nd}/2$) at very low concentrations, to that expected of a dimer ($t_{50}=t_{50}^\mathrm{nd}$) at very high concentrations.  \label{fig_sm_localeq}}
 \end{figure}

 \begin{figure}%[t]
 	\centering
 	\includegraphics[width=1\linewidth]{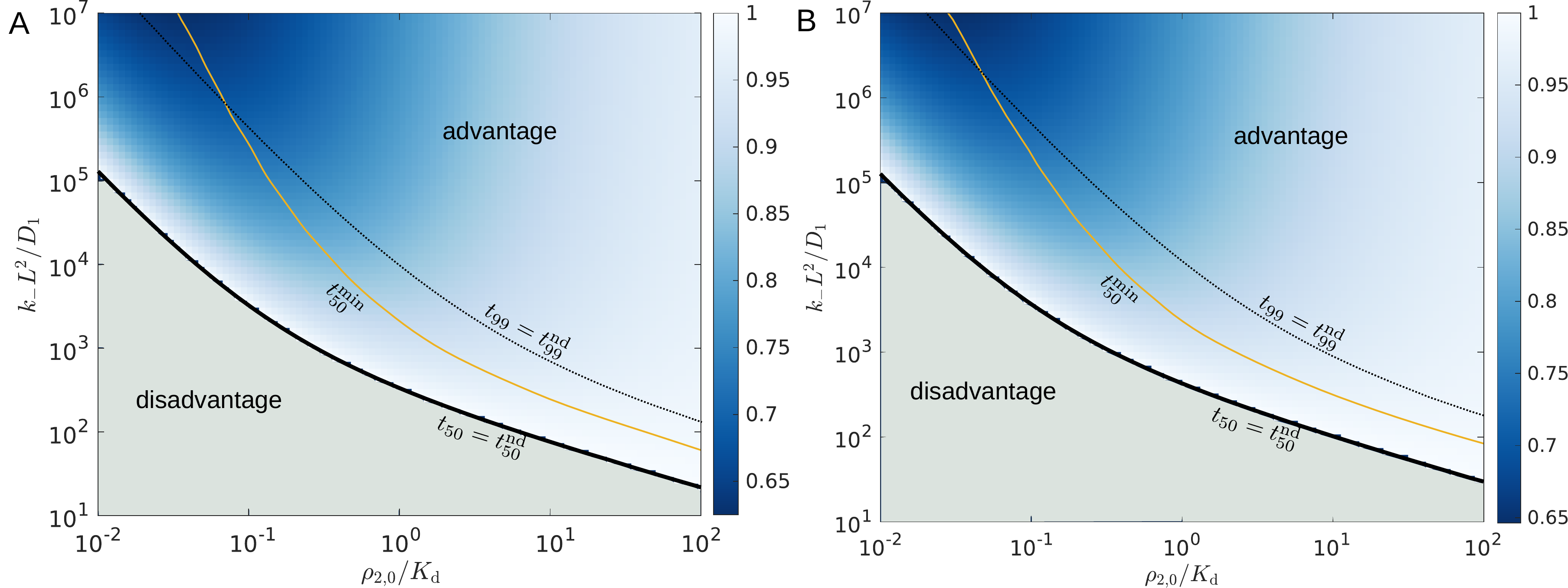}
 	\caption{Cooperatively enhanced reactivity of dissociating dimers with $D_2=0.5D_1$ in higher dimensions: (A) 2-D circular geometry and (B) 3-D spherical geometry.  $t_{50}/t_{50}^\mathrm{nd}$ as a function of both protein concentration and association-dissociation rate. The `disadvantage' region corresponds to $t_{50}/t_{50}^\mathrm{nd}>1$; the line labeled $t_{50}^\mathrm{min}$ indicates the concentrations that minimize $t_{50}$ for a given $k_-$; the line $t_{99}=t_{99}^\mathrm{nd}$ denotes the values above which $99\%$ of the proteins react faster in the dissociating case. Comparing to Fig.~2 in the main text, we see that the behavior is qualitatively and quantitatively nearly identical to the 1-D case, with advantages again as high as 40\%. \label{fig_sm_higherdim_tmedian}}
 \end{figure}

 \begin{figure}%[t]
 	\centering
 	\includegraphics[width=0.7\linewidth]{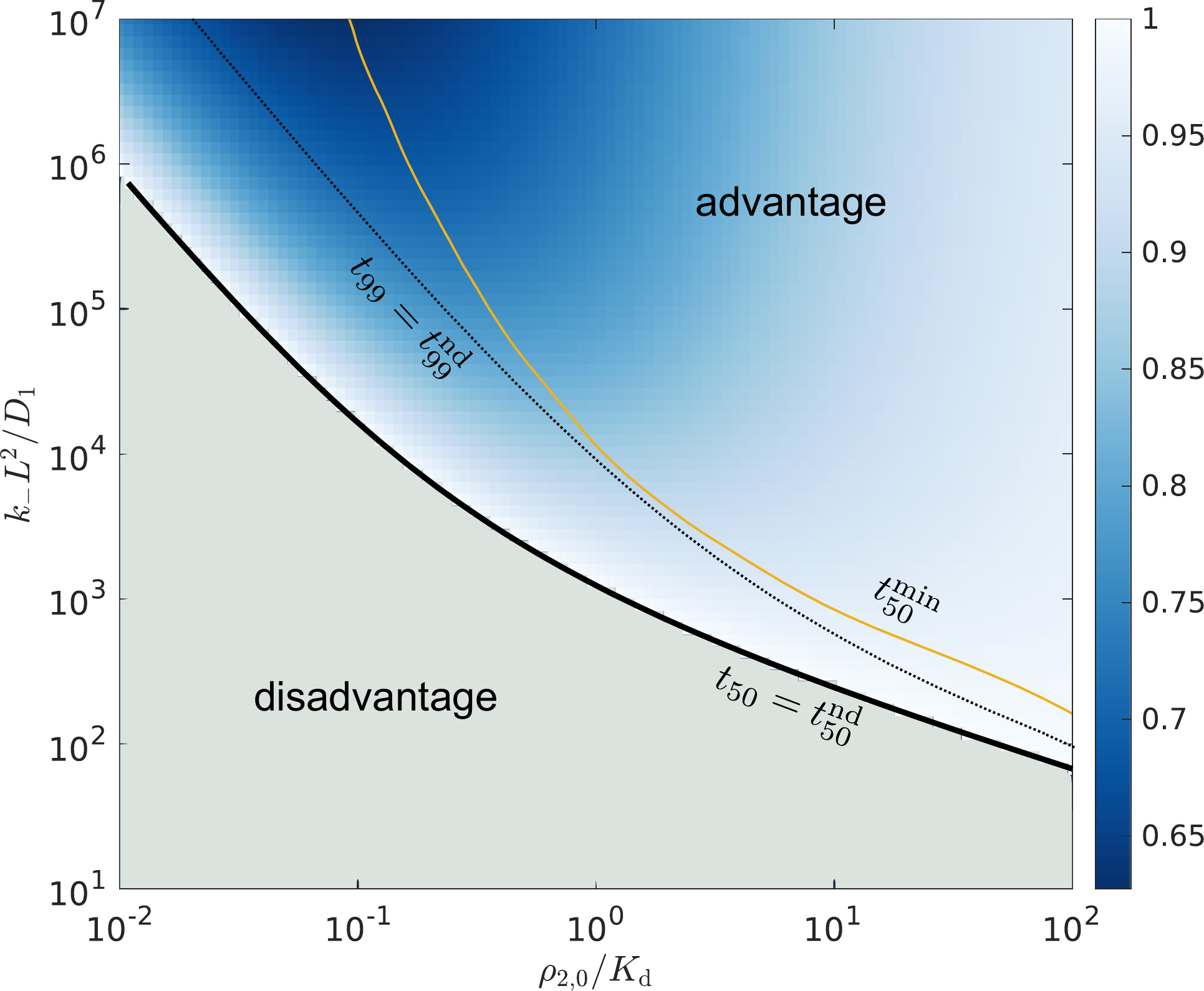}
 	\caption{Cooperatively enhanced reactivity of dissociating dimers with $D_2=0.5D_1$ starting from laterally uniform initial conditions in monomer-dimer binding equilibrium, i.e. satisfying $\rho_2 = \rho_1^2/\Kd$. This represents the case in which the solution is allowed to reach equilibrium in the absence of any reaction at the boundaries, and the reaction is then switched on at $t=0$. $t_{50}/t_{50}^\mathrm{nd}$ as a function of both protein concentration and association-dissociation rate. The `disadvantage' region corresponds to $t_{50}/t_{50}^\mathrm{nd}>1$; the line labeled $t_{50}^\mathrm{min}$ indicates the concentrations that minimize $t_{50}$ for a given $k_-$; the line $t_{99}=t_{99}^\mathrm{nd}$ denotes the values above which $99\%$ of the proteins react faster in the dissociating case. Comparing to Fig.~2 in the main text, we see that the behavior is qualitatively and quantitatively nearly identical to the case that starts with high protein concentration at the center, with advantages as high as 40\%.  \label{fig_sm_0p5_uniform}}
 \end{figure}
 
 \begin{figure}%[t]
 	\centering
 	\includegraphics[width=1\linewidth]{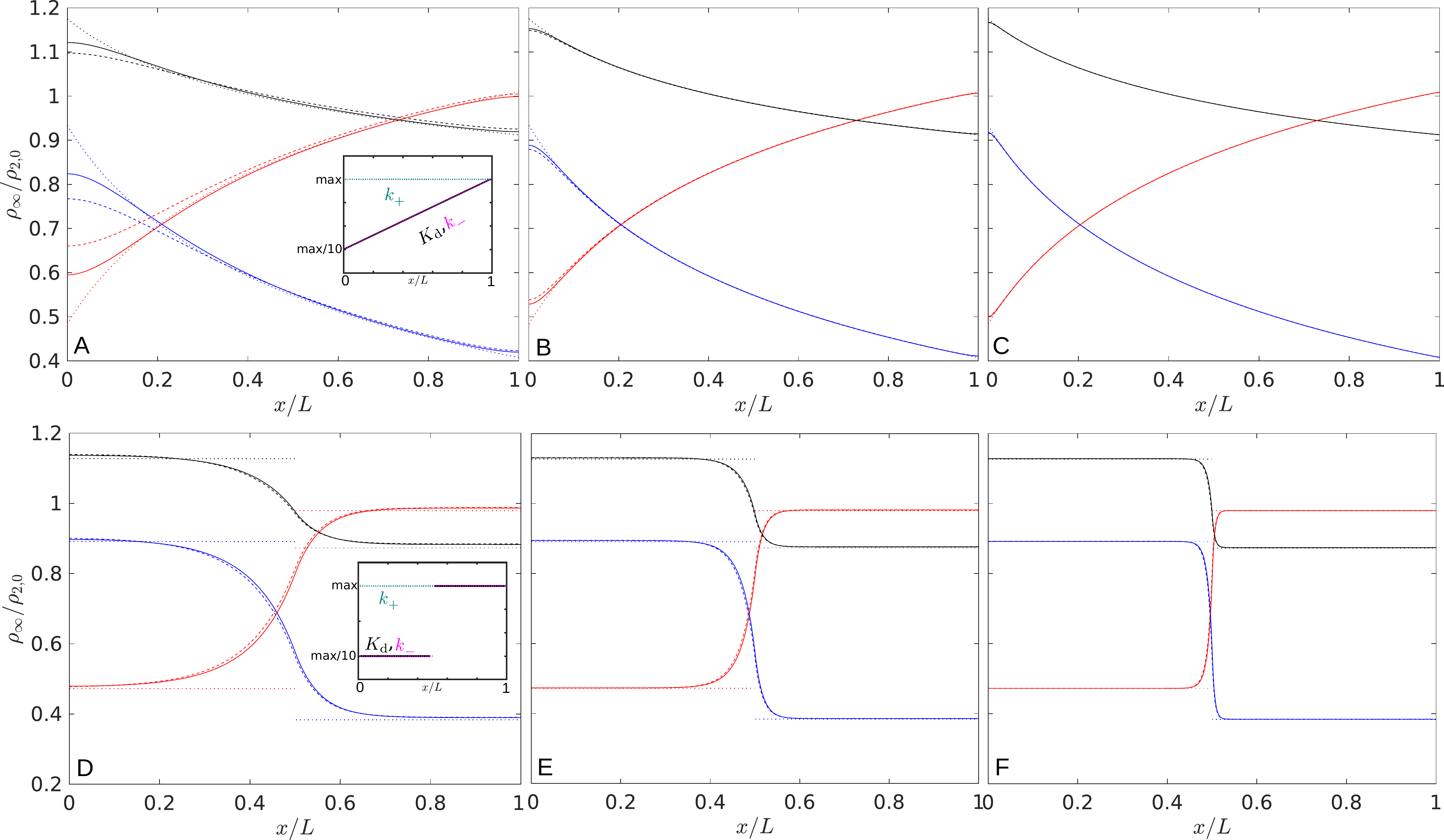}
 	\caption{Steady state profiles showing the increased accuracy of the local equilibrium approximation (dotted lines, Eq.~\textbf{7} in the main text) and the first order corrections (dashed lines, Eqs.~\ref{eq7} and \ref{eq8}) with increasing values of $k_- L^2 / D_1$. In (A--C) the increase in $\kmeff$ and thus $\Kd$ is linear, while in (D--F) it is step wise. From left to right, we use $k_- L^2 / D_1 = 10^2$ (A,D), $10^3$ (B,E), and $10^4$ (C,F). In all cases, we use $\rho_{2,0} = 0.4 \Kd^\mathrm{max}$. Note how the local equilibrium approximation (dotted lines) is independent of $k_- L^2 / D_1$, and how the first order approximation (dashed lines) correctly captures the decreasing width of the boundary regions with increasing $k_- L^2 / D_1$.     \label{fig_sm_corrections}}
 \end{figure}
 
 \begin{figure}%[t]
 	\centering
 	\includegraphics[width=1\linewidth]{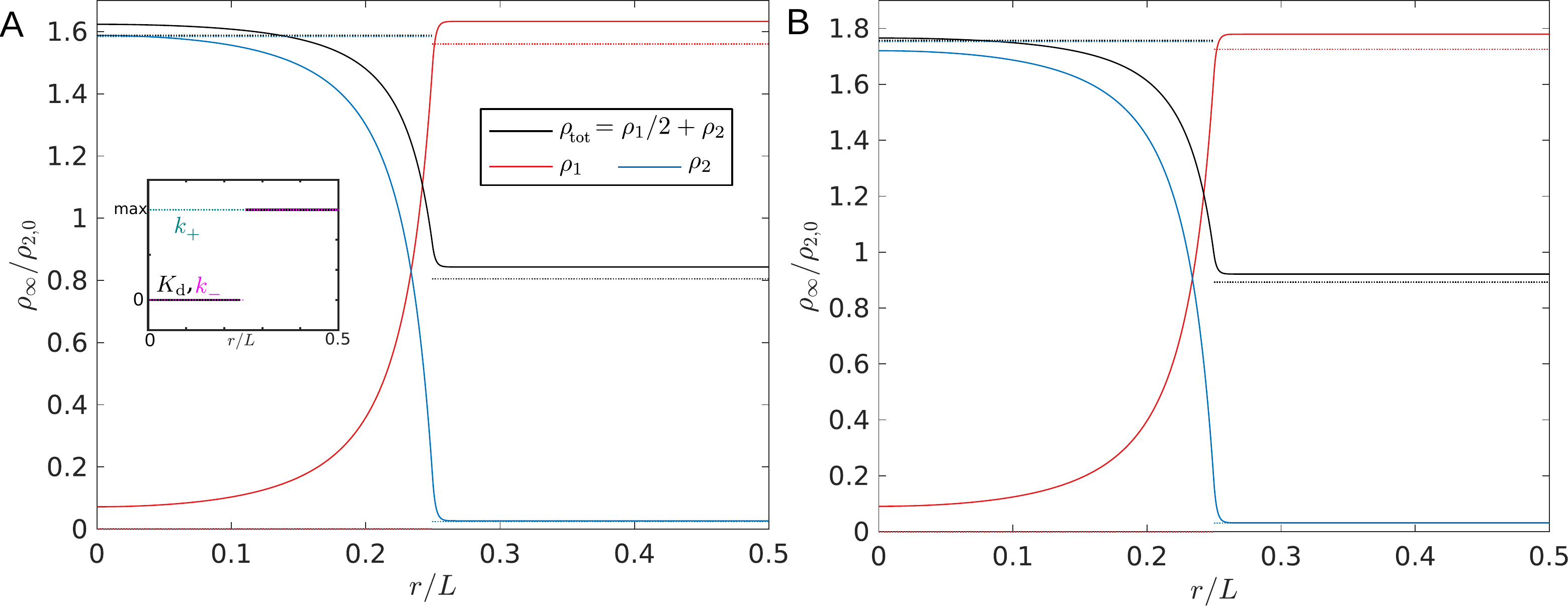}
 	\caption{Stabilitaxis in the presence of a dissociation gradient, as given by a discrete jump in $\kmeff$ and thus $\Kd$, in higher dimensions: (A) 2-D circular geometry and (B) 3-D spherical geometry. The solid lines are the steady state profiles resulting from numerical solution of the full equations, Eq.~[1] in the main text, whereas the dotted lines are the prediction from the local equilibrium approximation, Eq.~\textbf{7} in the main text. Parameters used are $D_2=0.5D_1$, $k_-^\mathrm{max} L^2 / D_1 = 10^5$, and $\rho_{2,0}=10^{-2} K_\mathrm{d}^\mathrm{max}$. Here, $L$ is the diameter of the circular (A) or spherical (B) domain, and $r$ is the radial coordinate.  \label{fig_sm_higherdim_grad}}
 \end{figure}
 
 \begin{figure}%[t]
 	\centering
 	\includegraphics[width=0.8\linewidth]{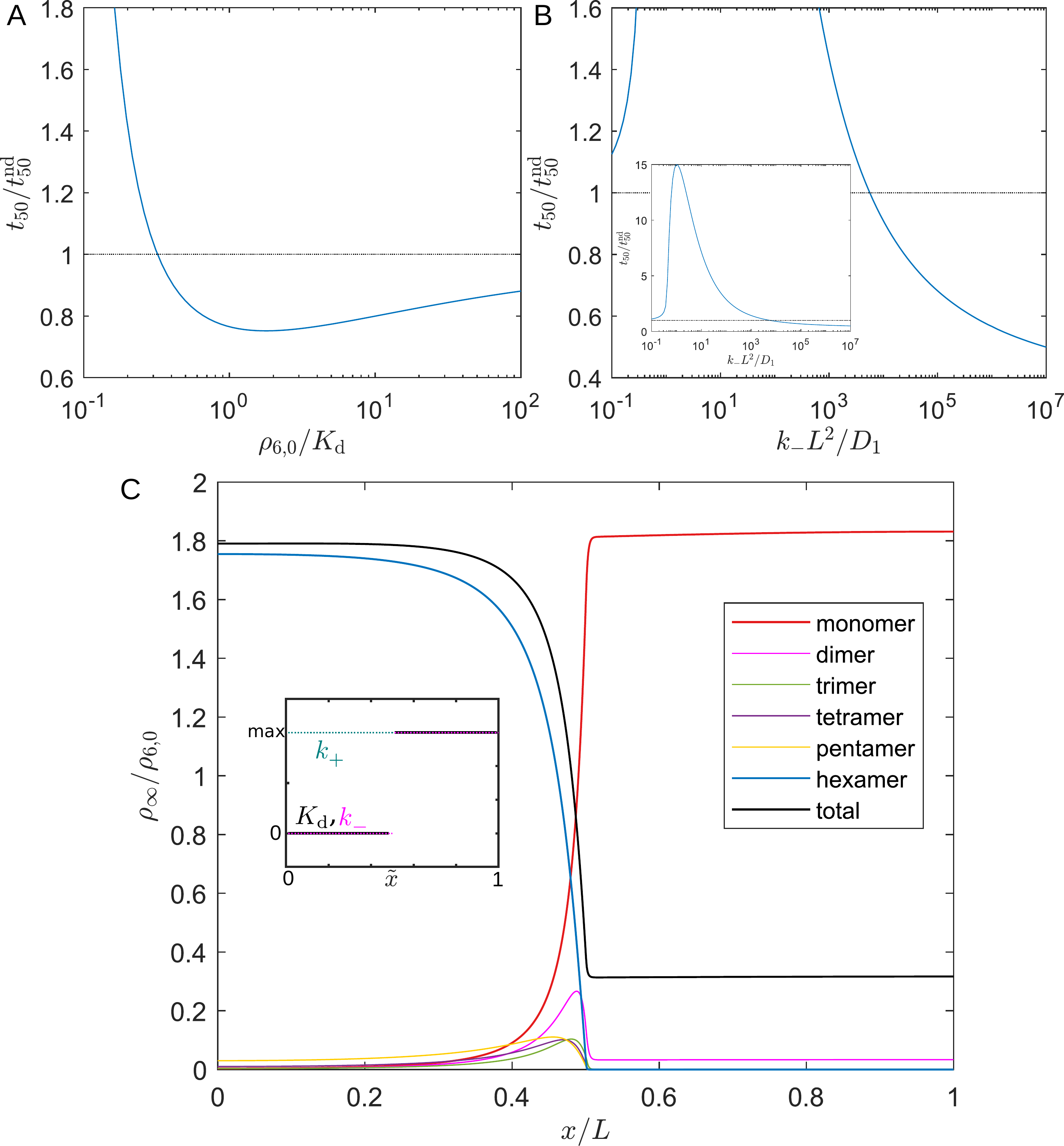}
 	\caption{Enhanced reactivity and stabilitaxis for a dissociating hexamer.  (A) Median first passage time $t_{50}$ relative to that of a non-dissociating protein $t_{50}^\mathrm{nd}$ as a function of protein concentration, and (B) as a function of $k_-  L^2/D_1$. The inset is a zoomed out version of the figure, showing that at slow association-dissociation rates reactivity is strongly slowed down. (C) Stabilitaxis in the presence of a dissociation gradient, as given by a discrete jump in $\kmeff$ and thus $\Kd$. The protein accumulates in the region where the hexamer form is stable, with six times higher concentration than in the high-dissociation region where the monomer form is preferred, as suggested by a generalization of Eq.~\textbf{8}. Parameters used are $D_n=D_1/n$ in all cases; $k_- L^2 / D_1 = 10^4$ in (A); $\rho_{6,0}=0.4K_\mathrm{d}$ in (B); $k_-^\mathrm{max} L^2 / D_1 = 10^5$ and $\rho_{6,0}=10^{-2} K_\mathrm{d}^\mathrm{max}$ in (C). \label{fig_sm_hexamer}}
 \end{figure}

\end{document}